\documentclass[prb,reprint,a4paper,superscriptaddress,showpacs]{revtex4-1}	
\relax
\usepackage{graphicx}
\usepackage{epsfig}
\usepackage{epstopdf}
\usepackage{amsmath}
\usepackage{amssymb}
\usepackage{hyperref}
\usepackage{dcolumn}	

\begin{document}
\title{Reply to ``Comment on `No evidence for orbital loop currents in charge-ordered  YBa$_2$Cu$_3$O$_{6+x}$ from polarized neutron diffraction' ''}	
\author{T. P. Croft}
\affiliation{H. H. Wills Physics Laboratory, University of Bristol, Bristol, BS8 1TL, United Kingdom.}

\author{E. Blackburn}
\affiliation{School of Physics \& Astronomy, University of Birmingham, Birmingham B15 2TT, United Kingdom.}
\author{J. Kulda}
\affiliation{Institut Laue-Langevin, 6, rue Jules Horowitz, BP 156, 38042 Grenoble Cedex 9, France.}
\author{Ruixing Liang}
\affiliation{Department of Physics \& Astronomy, University of British Columbia, Vancouver, British Columbia V6T 1Z1, Canada.}
\affiliation{Canadian Institute for Advanced Research, Toronto, Ontario M5G 1Z8, Canada.}
\author{D. A. Bonn}
\affiliation{Department of Physics \& Astronomy, University of British Columbia, Vancouver, British Columbia V6T 1Z1, Canada.}
\affiliation{Canadian Institute for Advanced Research, Toronto, Ontario M5G 1Z8, Canada.}
\author{W. N. Hardy}
\affiliation{Department of Physics \& Astronomy, University of British Columbia, Vancouver, British Columbia V6T 1Z1, Canada.}
\affiliation{Canadian Institute for Advanced Research, Toronto, Ontario M5G 1Z8, Canada.}
\author{S. M. Hayden}
\email{s.hayden@bris.ac.uk}
\affiliation{H. H. Wills Physics Laboratory, University of Bristol, Bristol, BS8 1TL, United Kingdom.}

\begin{abstract}	
The issues raised in the preceding comment of Bourges \textit{et al.} [\href{https://arxiv.org/abs/1710.08173}{arXiv:1710.08173}, \href{https://journals.aps.org/prb/abstract/10.1103/PhysRevB.98.016501}{Phys. Rev. B 98, 016501 (2018)}] are shown to be unfounded.  We highlight the complications caused by inhomogeneous beam polarization that can occur when using polarized neutron diffraction to detect small magnetic moments.

\end{abstract}
\pacs{74.72.-h, 74.25.Dw, 74.25.Ha, 75.25.+z}	
\maketitle
\section{Introduction}	

According to Varma's conjecture \cite{Varma1997_Varm,Varma2006_Varm} cuprate superconductors might exhibit a continuous phase transition to a phase which has circulating currents (CC) within each unit cell i.e. intra-unit cell (IUC) magnetic order.  This could then be observed by diffracting neutrons with their spins polarized parallel to the scattering vector $\mathbf{Q}$ and measuring the spin-flip (SF), $\sigma_{\uparrow\downarrow}$, and non-spin-flip (NSF), $\sigma_{\uparrow\uparrow}$, cross-sections. The coherent cross-sections for Bragg scattering are,
\begin{eqnarray}
\sigma_{\uparrow\downarrow}^{\textrm{Bragg}}&=&\left( \frac{d \sigma}{d \Omega}\right)_{\uparrow\downarrow}^{\mathbf{P} \parallel \mathbf{Q}}   =  \left( \frac{\gamma r_0}{2 \mu_B} \right)^2 \left|\mathbf{M}_{\perp}(\mathbf{G}) \right|^2 , 
\label{Eqn:xsct_mag} \\
\sigma_{\uparrow\uparrow}^{\textrm{Bragg}}&=&\sigma_{\downarrow\downarrow}^{\textrm{Bragg}}=\left( \frac{d \sigma}{d \Omega}\right)_{\uparrow\uparrow}^{\mathbf{P} \parallel \mathbf{Q}} =  \left| F_N (\textbf{G}) \right|^2 , 
\label{Eqn:xsct_nuc}
\end{eqnarray}
where $\left|\mathbf{M}_{\perp}(\mathbf{G}) \right|^2$ is the magnetic structure factor and $\left| F_N (\textbf{G}) \right|^2$ is the structure factor for nuclear Bragg scattering. $\left| F_N (\textbf{G}) \right|^2$ and $\left|\mathbf{M}_{\perp}(\mathbf{G}) \right|^2$ each include a Debye-Waller (DW) factor $\exp[-2M(T,Q)]$, which is temperature ($T$) and $\left| \mathbf{Q} \right|$ dependent (see Eq.~10-11 of Ref.~\onlinecite{Croft2017_CBKL_with_note}, also Eq.~4.50 and 4.57 of Ref.~\onlinecite{Lovesey1984_Love}). In addition to Bragg scattering, there may also appear diffuse nuclear coherent and (relatively weak) incoherent scattering whose SF and NSF components will give rise to background.  The $T$-dependence of all nuclear scattering follows the DW factor (see Eq.~4.52 of Ref.~\onlinecite{Lovesey1984_Love}).

Hence in an ideal experiment the $T$-dependence of the total NSF scattering $\sigma_{\uparrow\uparrow}^{\textrm{tot}}=\sigma_{\uparrow\uparrow}^{\textrm{Bragg}}+\sigma_{\uparrow\uparrow}^{\textrm{inc}}$ is determined by the DW factor with the scattered intensity increasing as $T$ is lowered. The rate of increase will vary between Bragg peaks and be greatest at large $\left| \mathbf{Q} \right|$.   In the absence of magnetic order (i.e. $\left|\mathbf{M}_{\perp}(\mathbf{G}) \right|^2 =0$), $\sigma_{\uparrow\downarrow}^{\textrm{tot}}$ will show the same DW factor. The onset of order introduces an additional $T$-dependence to $\sigma_{\uparrow\downarrow}^{\textrm{tot}}$.    In principle, we only need to determine the flipping ratio $R=\sigma_{\uparrow\uparrow}^{\textrm{Bragg}}/\sigma_{\uparrow\downarrow}^{\textrm{Bragg}}$. A knowledge of the crystal structure allows $\left| F_N (\textbf{G}) \right|^2$ to be calculated and hence the magnitude of $\sigma_{\uparrow\downarrow}^{\textrm{Bragg}}$ or $\left|\mathbf{M}_{\perp}(\mathbf{G}) \right|^2$ can be determined. 

In a practical realization of the above method, a beam of spin polarized neutrons is prepared by a polarizer, these neutrons are then scattered by the sample and the spin-state of the scattered neutrons is determined by an analyzer.  Real devices used to prepare and analyze the neutrons include Heusler crystal polarizers and benders based on supermirrors.  The beams produced by such devices are not 100\% polarized. In fact, they produce beams of spin-up and spin-down neutrons which may have somewhat different angular and spatial distributions. 
  
\begin{figure}[tb]
\includegraphics[width=0.95\linewidth]{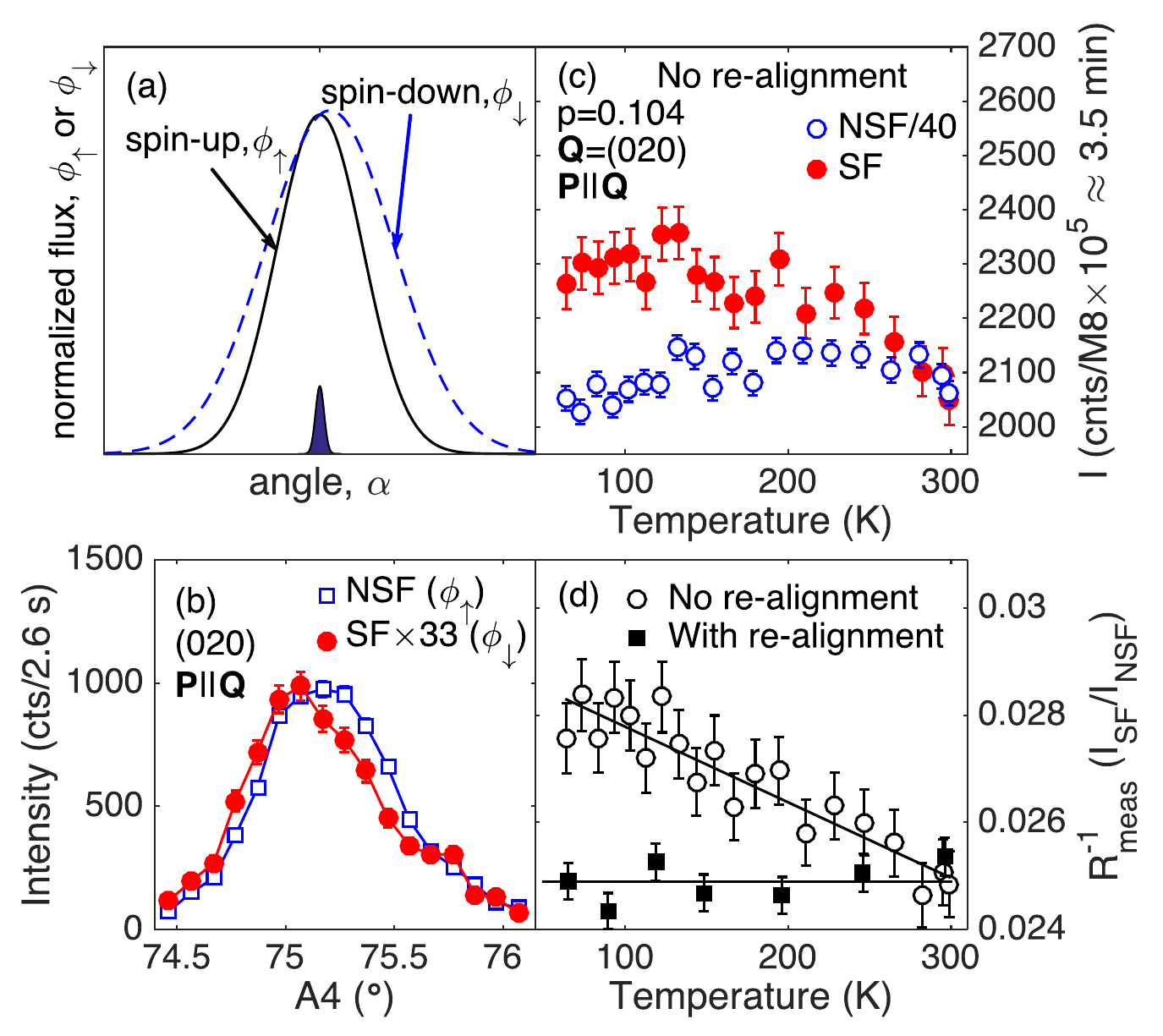}
\caption{ (a) Schematic angular flux distributions produced by a polarizing device, normalized by max[$\phi_{i}(\alpha)]$. The flux of the nominal polarization ($\phi_{\uparrow}$) is much larger than the other ($\phi_{\downarrow}$). The flux distributions are sampled by Bragg scattering as illustrated by shaded sample mosaic. (b) Measured flux distributions, $\phi_{\uparrow}$ and $\phi_{\downarrow}$ (from Fig.~6 of Ref.~\onlinecite{Croft2017_CBKL_with_note}) scaled to show the differences in form.   (c)  SF and NSF intensities of the (020) Bragg peak collected in a ``continuous temperature run'' in which the spectrometer was \textit{not moved}. (d) $R^{-1}(T)$ determined from the data in (c) (``No realignment'') (open circles). The spurious $T$-dependence observed can be removed (closed squares) by using a ``realignment'' measurement protocol where corrections are made for changing lattice parameter and sample movement (``With realignment''). }
\label{effect_of_correction}
\end{figure}

To illustrate the issues posed by this complication, consider the case of an imperfect polarizing system and a large ideal analyzer (to detect spin-up neutrons) used to measure a small sample which only exhibits nuclear coherent Bragg scattering (i.e. $\sigma_{\uparrow\uparrow}=\sigma_{\downarrow\downarrow} \ne 0, \sigma_{\uparrow\downarrow}=\sigma_{\downarrow\uparrow}=0$).  The polarizer will produce beams of neutrons with flux distributions $\phi_{\uparrow}(\alpha)$ and $\phi_{\downarrow}(\alpha)$, as seen from the sample position, which depend on the direction of travel of the neutron as specified by an angle $\alpha$ with respect to a reference direction at the sample position.  The distributions $\phi_{\uparrow}(\alpha)$ and $\phi_{\downarrow}(\alpha)$ may have different widths, skewness and centers as illustrated in Fig.~\ref{effect_of_correction}(a).  If the sample has a mosaic that is much less than the widths of distributions $\phi_{\uparrow}(\alpha)$ and $\phi_{\uparrow}(\alpha)$, it will Bragg scatter neutrons corresponding to a particular $\alpha$. By inserting an ideal spin flipper, we can choose the polarization of the detected neutron and measure a flipping ratio $R=\phi_{\uparrow}(\omega)/\phi_{\downarrow}(\omega)$ as a function of sample rotation $\omega$.  Small changes in $\omega$ cause corresponding changes in $\phi_{\uparrow}(\omega)$, $\phi_{\downarrow}(\omega)$ and $R$ (see Croft et al.\cite{Croft2017_CBKL_with_note} Fig.~6).  Similar considerations apply to a real analyzing device.  Thus, in a real experiment, the nominal $\sigma_{\uparrow \downarrow}$ channel contains not only the magnetic signal (Eqn.~\ref{Eqn:xsct_mag}), but also a spurious NSF nuclear Bragg scattering component, which is sensitive to temperature induced changes in sample position, orientation, lattice parameter and mosaic. In our search for orbital currents reported in Croft et al.\cite{Croft2017_CBKL_with_note}, we have minimized the effect of such $T$-induced changes in the measured $\sigma_{\uparrow \downarrow}$ by a specific measurement protocol.  
  
\section{Points Raised in the Comment}      

\subsection{Backgrounds and Magnetic Signal} 
Bourges et al. \cite{Bourges2018_BoSM} question our treatment of sample drifts with temperature. They claim that we are unable to determine the ``spin-flip reference line'' and that we do not have ``a proper knowledge of the background''.  As discussed in our original paper and above, our approach is to eliminate spurious $T$-dependent drifts in the signal and background. We argue that this is necessary because such changes (drifts) will: (i) be different in the SF and NSF channels, (ii) not necessarily be linear in $T$, and (iii) vary between Bragg peaks.  Our method consists in realigning the sample [$\omega$(A3), $2 \theta$(A4), sample tilts] to the maximum of $\phi_{\uparrow}(\alpha)$ at each temperature. Fig.~\ref{effect_of_correction} illustrates the effect of our protocol on the (020) peak. While a ``continuous temperature run'' method produces a linear variation of $R^{-1}(T)$, which is due mostly to a change in the scattering angle $2\theta$ caused by a change in the lattice parameter\cite{Bozin2016_BHSC}, the data collected using the ``realignment method'' produces a constant $R^{-1}(T)$. In contrast, Bourges et al. \cite{Bourges2018_BoSM} state that they observe $R^{-1}(T)$ which is ``linear in temperature with a positive slope'' at (020). 

An important point here is that data collected using our measurement protocol also directly reproduces the expected Debye-Waller $T$-dependence of the raw SF and NSF Bragg intensities (see Croft et al. \cite{Croft2017_CBKL_with_note} Figs.~8-9). For example, Eqn.~11 of Ref.~\onlinecite{Croft2017_CBKL_with_note} predicts that the intensity of the (011) and (020) peaks increase by $\sim 2$\% and $\sim 7$\% respectively between 300K and 60K. The $T$-dependence of the incoherent background [see e.g. Ref.~\onlinecite{Croft2017_CBKL_with_note} Fig.~7(c,d)], which also obeys the DW-factor, is also consistent with this calculation.   Our method\cite{Croft2017_CBKL_with_note} is fundamentally different to that discussed in Bourges et al. \cite{Bourges2018_BoSM} in that it removes the spurious $T$-dependence of the intensity in both the SF and NSF channels.

\subsection{Samples and Sample Size}
Our admittedly smaller sample size and small mosaic has advantages. The sample may be used to characterize the polarizer/analyzer system and find a consistent alignment of the experiment avoiding drifts with changing temperature. All of the sample is in the diffraction condition, yielding more scattered neutrons per unit of volume and permitting a consistent measurement of Bragg intensities. Ultimately, this is a counting experiment and the error bars (shown in our plots) are determined from the number of counts. 

The comment\cite{Bourges2018_BoSM} states that sample C ``was synthesized using the same self-flux method using a BaZrO$_3$ crucible''. This appears not to be the case.   It is stated in Fauqu\'{e} \textit{et al.} \cite{Fauque2006_FSHP} that sample C was studied by Hinkov \textit{et al.} \cite{Hinkov2004_HPBS}.  The authors\cite{Keimer2018_Keim} of that paper and the references therein indicate that other crucible types\cite{Lin1992_LZLS, Liang1992_LDBB} were used to produce sample C. 

\begin{figure}
\includegraphics[width=0.7\linewidth]{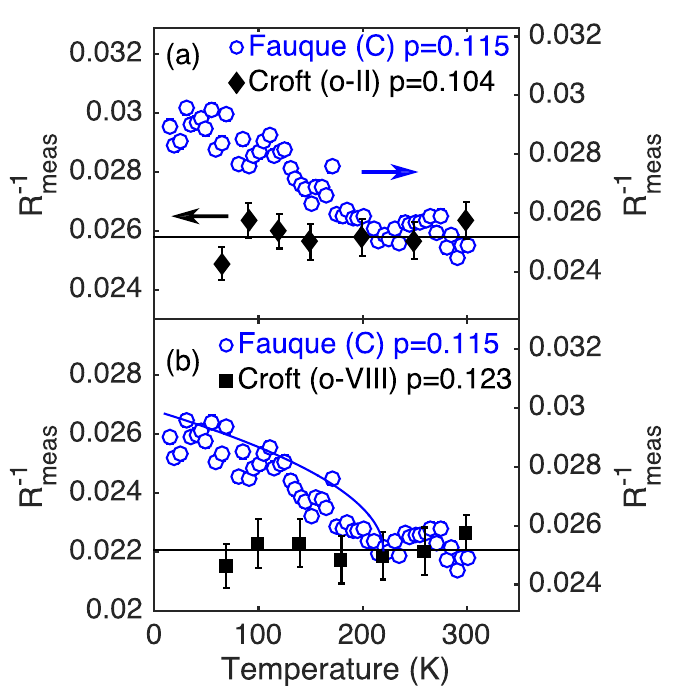}
\caption{Comparison of raw data. Inverse of the measured flipping ratio, $R^{-1}$, for detwinned underdoped YBa$_2$Cu$_3$O$_{6+x}$ samples at $\textbf{Q}=(011)$. (a) Black diamonds are data of Croft \textit{et al.} for ortho-II  YBa$_2$Cu$_3$O$_{6.54}$($p$=0.104). (b) Black squares are ortho-VIII  YBa$_2$Cu$_3$O$_{6.67}$($p$=0.123). In both cases, $R^{-1}$ is compared with that calculated from the raw data of Fauqu\'{e} \textit{et al.}\cite{Fauque2006_FSHP} Fig.~1(a), sample C (YBa$_2$Cu$_3$O$_{6.6}$, $p$=0.115). These data are shown as blue open circles.  Bourges et al. \cite{{Bourges2018_BoSM}} state that the raw data do not change appreciably with doping in the range  ($p=0.1-0.12$). Theory\cite{Fauque2006_FSHP,Croft2017_CBKL_with_note} (e.g. Croft \textit{et al.}\cite{Croft2017_CBKL_with_note} Eqn.~8) suggests that changes in $R^{-1}$ are comparable, so the offsets of the left and right axes are different to accommodate different residual instrumental flipping ratios $R^{-1}_{\mathrm{inst}}$.  The blue solid line in (b) is that shown schematically in Croft \textit{et al.}\cite{Croft2017_CBKL_with_note} Fig.~9(c). }
\label{raw_data_comparison}
\end{figure}
\subsection{Raw Data}
Bourges \textit{et al.}\cite{Bourges2018_BoSM} compare the raw data of the two experiments.  One should bear in mind that our and their experiments were carried out in different ways using different samples on different instruments. Different methods were used to obtain the final magnetic cross-sections. We believe the only meaningful way to do this is to compare the resulting cross-sections, which eliminate, as much as possible, the influences of diverse instrumental factors. Nevertheless, Bourges et al. \cite{Bourges2018_BoSM} state that the raw data do not change appreciably with doping in the range ($p=0.1-0.12$). In Fig.~\ref{raw_data_comparison} we superpose the raw data for their sample C ($p$=0.115)\cite{Fauque2006_FSHP} with the ortho-II and ortho-VIII samples of Croft \textit{et al.}\cite{Croft2017_CBKL_with_note}. See the caption of Fig.~\ref{raw_data_comparison} for details.

Fig.~2 of the comment\cite{Bourges2018_BoSM} reproduces and questions lines from Croft \textit{et al.}\cite{Croft2017_CBKL_with_note} Fig.~8). We note these are simply predictions of the changes in flipping ratio corresponding to the published $T$-dependent magnetic cross-sections in Fauqu\'{e} \textit{et al.}\cite{Fauque2006_FSHP} Figs. 1(c) and 2(b). Details of the calculation are given in the corresponding figure captions in Croft \textit{et al.} \cite{Croft2017_CBKL_with_note}. 

\subsection{Alternative Analysis}
In the comment section titled ``Alternative Analysis'' it is stated: ``The inverse of flipping ratio at the reflection (020) is linear in temperature with a positive slope as it has been shown to exist $\dots$, this slope is inevitable as the sample drifts in the neutron beam upon changing temperature. Croft \textit{et al.} arbitrarily describe it with a flat horizontal line only. At the accuracy required to observe the IUC magnetic signal, this is not a correct approximation.''  Bourges et al.\cite{Bourges2018_BoSM} then use our (020) data to obtain a ``gray shaded area'' shown in their Fig. 2.  This area represents the ``combined effects of the statistical errors of each point, occurrence of off-statistical points''. Our view is simply that there is no observable signal at (020).  The ``inevitable slope'' mentioned above is eliminated by our measurement protocol so that our $R^{-1}(T)$ measured for $p = 0.014$ and (020) can be expected to exhibit no $T$-dependence, which is statistically consistent (within a $\chi^2$ goodness-of-fit test) with our data. The statistical significance of a our data is encoded in the error bars shown and derived from Poisson statistics in the usual way.  It is not obviously described by the gray areas which are arbitrarily transferred between different experimental conditions and counting times. 

\subsection{Other experiments}
Our original motivation for the search for orbital magnetic order was to understand the nature of the broken symmetry in the pseudogap (PG) phase. Anomalies which may correspond to a broken symmetry have now been seen by macroscopic probes including the Kerr effect\cite{Xia2008_XSDK}, resonant ultrasound\cite{Shekhter2013_SRLH}, and optical second-harmonic generation\cite{Zhao2017_ZBLB}. Changes in anisotropy are also observed in thermopower\cite{Daou2010_DCLC} and susceptibility\cite{Sato2017_SKMK}.  However,  IUC magnetic order is a specific proposal\cite{Varma1997_Varm,Varma2006_Varm} which is not directly probed by such experiments.  

Bourges et al. \cite{Bourges2018_BoSM} say that we have not properly taken into account other experiments ``the comparison with local probes results in Ref.~\onlinecite{Croft2017_CBKL_with_note} is outdated and partial as it dismisses the recent literature\cite{Zhang2018_ZDTH} about muon spin relaxation ($\mu$SR) results.'' The search for loop currents with $\mu$SR has been controversial, and still appears that way.  Zhang \textit{et al.}\cite{Zhang2018_ZDTH} have recently studied more highly doped compositions (i.e. YBa$_2$Cu$_3$O$_{6.72}$ and above) than those investigated by Croft \textit{et al.} \cite{Croft2017_CBKL_with_note} and concluded there are slow magnetic fluctuations at low $T$.   However, Sonier \textit{et al.}\cite{Sonier2002_SBKH,Sonier2001_SBKM,Sonier2009_SPSH} have used $\mu$SR to study YBa$_2$Cu$_3$O$_{6.6}$ and YBa$_2$Cu$_3$O$_{6.67}$ and found no local fields of the size expected if loop-currents were present.  Another widely used local probe is NMR, where a recent study by Wu \textit{et al.}\cite{Wu2015_WMKH}, on a sample like those investigated in Croft  \textit{et al.} \cite{Croft2017_CBKL_with_note}, observed no evidence for loop currents.   

Bourges et al. \cite{Bourges2018_BoSM} also write in their supplementary material (SM), referring to Zhao \textit{et al.} \cite{Zhao2017_ZBLB}: ``second harmonic generation optical measurements found an odd-parity magnetic order parameter exactly in the same temperature and doping ranges, fully consistent with the loop current-type phase.'' However, it is stated in Zhao \textit{et al.} \cite{Zhao2017_ZBLB} that the second harmonic data are consistent with (i) magnetic point groups that break inversion and time-reversal symmetry and also with (ii) point groups that break inversion alone.  Thus this experiment is also consistent with the absence of loop-currents.

\subsection{Finite Correlation Length of Order}
Previous to their comment Bourges \textit{et al.} \cite{Bourges2018_BoSM} have stated that the underdoped composition exhibits a ``3D magnetic order'' \cite{Sidis2013_SiBo} giving rise to ``resolution limited'' diffraction maxima\cite{Fauque2006_FSHP,Mook2008_MSFB,Mangin-Thro2015_MSWB}. The comment authors wish to see the effect of a relatively short correlation length of $\xi_c=75$~\AA\ on our conclusions.  We agree that if the order had a finite $\xi_c$, it would reduce our (and their) signal. However we note: (i) their published work\cite{Mook2008_MSFB} is also consistent with an infinite $\xi_c$ for the dopings studied here, thus there is no evidence for a such a short $\xi_c$; (ii) our limits placed on the cross section due to 3D magnetic order remain unchanged.  Moreover, in our SM\cite{Croft2018_supp} we show that for $\xi_c=75$~\AA, the reduction factor would be 1.4 not 3.

\begin{figure}[tbh]
\includegraphics[width=0.8\linewidth]{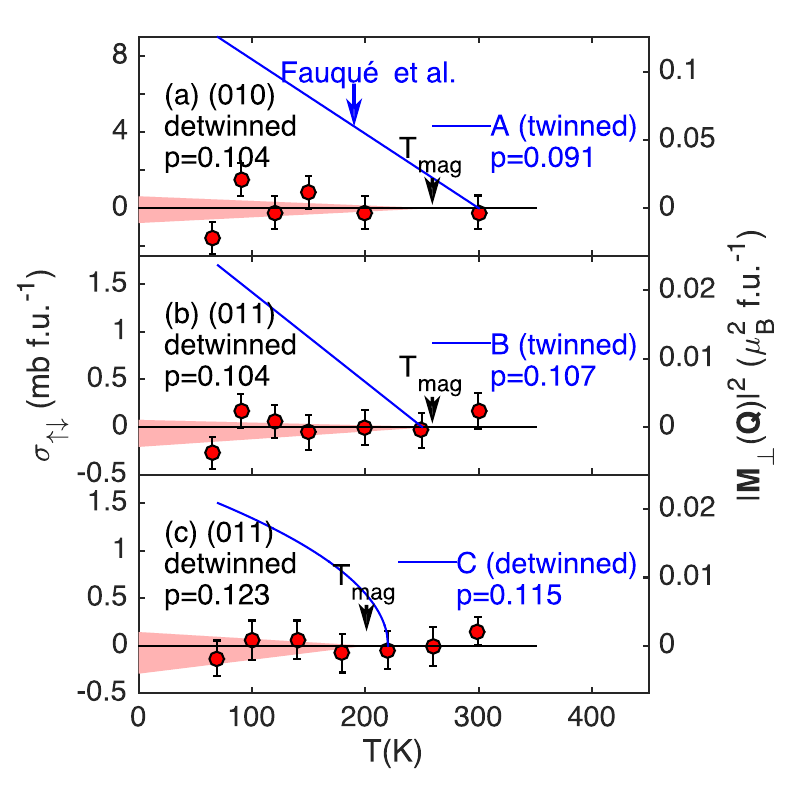}
\caption{$T$-dependence of the magnetic cross section $\sigma_{\uparrow\downarrow}$ and magnetic structure factor squared $\left|\mathbf{M}_{\perp}(\mathbf{G}) \right|^2$ determined using Croft \textit{et al.} \cite{Croft2017_CBKL_with_note} Eqns.~8 and 9 with values of $|F_N|^2_{\textrm{calc}}$ from Table~1 of our SM. The solid blue lines show indicative values of $\sigma_{\uparrow\downarrow}$ from Fauqu\'{e} \textit{et al.} \cite{Fauque2006_FSHP} Fig.~1(c) and 2(b). Note that these values have recently been revised downwards by 15-20\% in Ref.~\onlinecite{Bourges2018_BoSM} and that the comparison of twinned samples with detwinned samples requires averaging over domains.}
\label{magmo_Tdep}
\end{figure}  
   
\subsection{Nuclear Structure Factor.}
In various places in the comment it is claimed that our calibration procedure is incorrect by a factor of $\sim3$. It is argued that the predicted changes in $R^{-1}(T)$ are too high due to the value of $|F_N|^2$ used in Eq.~\ref{Eqn:xsct_nuc} of this reply and Croft \textit{et al.} Eqn.~6 being too small. It is further claimed that our consistency check on $|F_N|^2$ does not correctly take into account resolution effects and that we need to take into account the variation of $|F_N|^2$ with oxygen stoichiometry. These issues are addressed below.

\subsubsection{Oxygen Ordering and its effect on the Nuclear Structure Factor.} 
In Sec. II.D of the comment supplementary material\cite{Bourges2018_BoSM} (SM), it is claimed that Eqn.~12 of Croft \textit{et al.} \cite{Croft2017_CBKL_with_note} cannot be used to determine the $|F_N|^2$.  Specifically, it is stated that the equation: (i) ``assumes the oxygen chain site O(1) is randomly occupied'' and (ii) there is ``destructive interference'' between the normal neutron coherent structure factor and scattering function $S(Q)$ due to ``short range oxygen ordering''. Clearly, the oxygen chains are not fully ordered and this leads to diffuse scattering being observed by x-rays, neutrons and electron microscopy. However, statements (i) and (ii) appear to be incorrect. We can show this easily in one dimension: a proof is given in our\cite{Croft2018_supp} SM. The proof can be extended to higher dimensions and to include other atoms in the structure. 

\subsubsection{Data Calibration and Resolution Corrections}
In order to make our result more robust, we checked in Croft \textit{et al.}\cite{Croft2017_CBKL_with_note} Fig.~10 that the measured $|F_N|^2$ of four Bragg reflections where consistent with the structure of Jorgensen \textit{et al.} \cite{Jorgensen1990_JVPN}. We used an approximate ``Lorentz factor'' to correct our data for changing experimental resolution between reflections, which the comment claims is not appropriate. In our SM\cite{Croft2018_supp} we re-analyze our data using a full resolution calculation\cite{Popovici1975_Popo,ResCal2009,Zachariasen1967_Zach,Schultz1986_Schu} and show that under our experimental conditions, the Lorentz approximation is sufficient.

\subsubsection{Variation of structure factors with oxygen stoichiometry}
In the SM (Sec. II.C) of the comment it is pointed that $|F_N(\mathbf{G}=(011))|^2$ varies with doping. In our original analysis, we used a structure factor for $x=0.54$, based on the structure of Ref.~\onlinecite{Jorgensen1990_JVPN} placing the excess oxygen on the O1 sites, to normalize both samples investigated.  Using the actual oxygen stoichiometry yields $|F_N(\mathbf{G}=(011))|^2$ values of 0.28 and 0.20 barn f.u.$^{-1}$ for $x=0.54$ and 0.67 respectively.  Fig.~\ref{magmo_Tdep} shows our data analyzed with these values. This change has little effect on our conclusion.

\section{Conclusion}	
Based on the arguments given in the preceding sections we come to the conclusion that the objections raised by Bourges et al. \cite{{Bourges2018_BoSM}} are unfounded.  Fig.~\ref{magmo_Tdep} summarizes our main result, the estimation of the magnetic cross section in barns. Here we have used the new values of $|F_N(\mathbf{G}=(011))|^2$ which take account of the variation of oxygen content \cite{Bourges2018_BoSM}. Our conclusion remains unchanged: We find no evidence of a magnetic signal of the order of 1-2 mbarn~f.u.$^{-1}$ as was reported by Fauqu\'{e} \textit{et al.} \cite{Fauque2006_FSHP} and subsequent work \cite{Fauque2006_FSHP,Mook2008_MSFB,Sidis2013_SiBo,Mangin-Thro2015_MSWB}. 
In our view, the way forward is to re-assess the previous work in terms of the two beam picture (combined with the $T$-dependence of sample position, lattice parameter, orientation and mosaic) which we have presented in our original paper and this reply. The common feature of our work and the work of Bourges and collaborators is that we make use of polarized mirror benders or Heusler crystal polarizers to polarize and analyze the neutron spin state.  The production of spin-up and spin-down beams which have different angular and spatial distribution functions is inevitable at some level when these devices are used. This should be explicitly addressed in designing the experimental method.

\section{Acknowledgements}
This work was supported by the UK EPSRC (Grant EP/R011141/1).

%

\end{document}


\title{No Evidence for Orbital Loop Currents in Charge Ordered YBa$_2$Cu$_3$O$_{6+x}$ from Polarized Neutron Diffraction: Reply to Comment, Supplementary Material}	
\author{T. P. Croft}
\affiliation{H. H. Wills Physics Laboratory, University of Bristol, Bristol, BS8 1TL, United Kingdom.}

\author{E. Blackburn}
\affiliation{School of Physics \& Astronomy, University of Birmingham, Birmingham B15 2TT, United Kingdom.}
\author{J. Kulda}
\affiliation{Institut Laue-Langevin, 6, rue Jules Horowitz, BP 156, 38042 Grenoble Cedex 9, France.}
\author{Ruixing Liang}
\affiliation{Department of Physics \& Astronomy, University of British Columbia, Vancouver, Canada.}
\affiliation{Canadian Institute for Advanced Research, Toronto, Canada.}
\author{D. A. Bonn}
\affiliation{Department of Physics \& Astronomy, University of British Columbia, Vancouver, Canada.}
\affiliation{Canadian Institute for Advanced Research, Toronto, Canada.}
\author{W. N. Hardy}
\affiliation{Department of Physics \& Astronomy, University of British Columbia, Vancouver, Canada.}
\affiliation{Canadian Institute for Advanced Research, Toronto, Canada.}
\author{S. M. Hayden}
\email{s.hayden@bris.ac.uk}
\affiliation{H. H. Wills Physics Laboratory, University of Bristol, Bristol, BS8 1TL, United Kingdom.}

\begin{abstract}	
Further detail is provided in support of our reply to the comment of Bourges \textit{et al.}\cite{Bourges2018_BoSM} on Croft \textit{et al.}\cite{Croft2017_CBKL}.   

\end{abstract}
\pacs{74.72.-h, 74.25.Dw, 74.25.Ha, 75.25.+z}	
\maketitle
\subsection{Evidence of Intra Unit Cell Magnetic Order from Polarized Neutron Diffraction}	

Bourges et al. \cite{Bourges2018_BoSM} point out that there are an extensive series of papers in which polarized neutron diffraction has been used to study intra unit cell (IUC) magnetic order. These studies have been performed using several different spectrometers on samples of different cuprate systems prepared by different methods. We have carried out one experiment using a different experimental method which yielded a null result\cite{Croft2017_CBKL}. In our view, the way forward is to re-assess the previous work in terms of the ``two beam'' picture which we have presented in our original paper and the main body of reply. The analysis should take into account the effects of the $T$-dependence of sample position, lattice parameters, orientation and mosaic. The common feature of our work and the work of Bourges and collaborators is that we make use of polarized mirror benders or Heusler crystal polarizers to polarize and analyze the neutron spin state.  The production of spin-up and spin-down beams which have different angular and spatial distribution functions is inevitable at some level when these devices are used. Guide field spatial inhomogeneities will also contribute, creating a depolarized beam with a different distribution to the original polarized one. These effects can be minimized (as in our protocol). We note that the use of ${}^{3}$He polarizers may offer some advantages for future work. 

Our experiment was carried out with the applied magnetic field at the sample position parallel to the scattering wavevector. It is possible to vary the direction of this field with respect to the scattering vector.  This yields further information, as the moments detected in spin-flip scattering are perpendicular to the applied field. However, changing the field direction may also change the feedthrough between the spin flip and non spin flip channels creating further complications in interpretation when a Bragg peak is present in one of the channels.

\subsection{Finite Correlation Length of Order}

In the case of finite $\xi_c$=75\;\AA, we describe the correlations with a scattering function $S(q_z) \propto (\xi_c/\sqrt{2 \pi}) \exp(-q_z^2 \xi_c^2/2)$. The Gaussian form is used for ease of  integration and we assume $\delta$-functions for $q_x,q_y,\hbar\omega$.  The measured intensity $I$ is then obtained by multiplying $S(q_z)$ by the instrumental resolution function $R_z \exp (-q_z^2/2 \sigma^2_{\textrm{res}})$ and integrating over $q_z$. Evaluating this integral, we find $I \propto  \xi_c /\sqrt{\xi_c^2+\sigma_{\textrm{res}}^{-2}}$ i.e. the effect of a finite $\xi_c$ is to reduce $I$ by this factor with respect to the case of infinite $\xi_c$.  Our instrumental resolution determined from A3 scans (see captions\cite{Croft2017_CBKL} Figs.~6 and 7) is $0.57^{\circ}$, hence $\sigma_{\textrm{res}} \approx 0.0076$~\AA$^{-1}$, yielding a reduction factor of $0.49$. For the case of an A3 width of $1^{\circ}$ quoted in the comment, we have $\sigma_{\textrm{res}} \approx 0.013$~\AA$^{-1}$ yielding 0.70.   Hence the ratio of the relative sensitivities would be 1.4 not 3.   

\subsection{Oxygen Ordering and its effect on the Nuclear Structure Factor.} 

In Sec.~II.D of the comment supplementary material\cite{Bourges2018_BoSM}, it is claimed that Eqn.~12 of Croft \textit{et al.} \cite{Croft2017_CBKL} cannot be used to determine the structure factors of our nuclear Bragg peaks.  Specifically, it is stated that the equation: (i) ``assumes the oxygen chain site O(1) is randomly occupied'' and (ii) there is ``destructive interference'' between the normal neutron coherent structure factor and scattering function $S(Q)$ due to ``short range oxygen ordering''. Here we show these statements are incorrect in one dimension.  

Consider a 1-D chain with sites at positions $x_k=ak$, where $k$ is an integer and $a$ the lattice parameter. Let the occupancy of the site $k$ be $n_k$. The number density is
\begin{equation}
\rho(x) \equiv \sum_k n_k \, \delta(x-x_k).
\end{equation}
The two-point density-density correlation function is:
\begin{equation}
C_{\rho\rho} = \langle \rho(x_1) \, \rho(x_2) \rangle.
\end{equation} 
The structure factor for scattering is the Fourier transform of this:
\begin{eqnarray}
\frac{d\sigma}{d \Omega} & = & \int \int e^{iq(x_1-x_2)} \langle \rho(x_1) \, \rho(x_2) \rangle \, dx_1 \, dx_2 \\
&=& \sum_{kk^{\prime}} e^{iq (x_k-x_{k^{\prime}})} n_k n_{k^{\prime}}.
\end{eqnarray}
We can split the site occupancy into an average value and fluctuations and evaluate the scattering cross section:
\begin{equation}
n_k= \langle n \rangle+ \Delta n_k
\end{equation}
\begin{eqnarray}
\frac{d\sigma}{d \Omega} & = & \langle n \rangle^2 \sum_{kk^{\prime}} e^{iq (x_k-x_{k^{\prime}})}  
+ 2 \langle n \rangle \sum_{kk^{\prime}} e^{iq (x_k-x_{k^{\prime}})} \Delta n_k \nonumber \\
&+& \sum_{kk^{\prime}} e^{iq (x_k-x_{k^{\prime}})}  \Delta n_k \Delta n_{k^{\prime}}, \label{n_expand} \\
&=& 2 \pi \langle n \rangle^2 \sum_m \delta \left( q-m \frac{2 \pi}{a} \right)+ S_{\rho\rho}(q), \label{Eqn:S_qw}
\end{eqnarray}
where $m$ is an integer. The first term in Eqn.~\ref{Eqn:S_qw} corresponds to Bragg scattering and the second to diffuse scattering due to correlations in the site occupancy. The second term in Eqn.~\ref{n_expand} is zero at a Bragg position because $\langle \Delta n_k \rangle=0$. Note that: (i) we have made no assumption of random site occupancy (correlations are allowed). (ii) There is no interference term between the diffuse scattering $S_{\rho\rho}(q)$ and the Bragg scattering.  The proof can be extended to higher dimensions and to include other atoms in the structure.
\vspace{5mm}

\subsection{Data Calibration and Resolution Corrections}
Our use of a Lorentz factor to describe the changing instrumental resolution between Bragg peaks is questioned in Sec.~V of the supplementary material of the comment\cite{Bourges2018_BoSM}. Here we compute the effect of the changing instrumental resolution on our diffraction data, using the Popovici method \cite{Popovici1975_Popo,ResCal2009} applicable to three-axis spectrometers. The measured  neutron scattering cross section is related to the scattering function $S(\mathbf{Q},\omega)$ though the equation:
\begin{multline}
\frac{d^2 \sigma}{d\Omega \, dE} \propto R_0(\boldsymbol{\mathcal{Q}}_0) \int d^4  \mathcal{Q} \; S(\boldsymbol{\mathcal{Q}}) \times \\
\exp \left[ -\frac{1}{2} \sum_{ij}(\mathcal{Q}_i-\mathcal{Q}_i^0)  M_{ij}(\boldsymbol{\mathcal{Q}}^0) (\mathcal{Q}_j-\mathcal{Q}_j^0) \right], \label{res_convolution}
\end{multline} 
where $\boldsymbol{\mathcal{Q}} \equiv (Q_x, Q_y, Q_z, \hbar \omega)$ describes the momentum and energy transfer is a 4-D space and $\boldsymbol{\mathcal{Q}}^0$ is the nominal spectrometer setting. Including the sample mosaic in the resolution function and assuming elastic Bragg scattering, we are able to compute the resolution correction to the integrated intensity measured in an A3 ($\omega$) Bragg scan. This is known as the Lorentz factor, $1/L$. In our original paper, we made the approximation $1/L \approx \sin(2\theta)$.  This approximation worked because the Soller collimators were removed and the collimation was defined by apertures.  The sample also had a small mosaic. 

\begin{figure}[htb]
\includegraphics[width=0.8\linewidth]{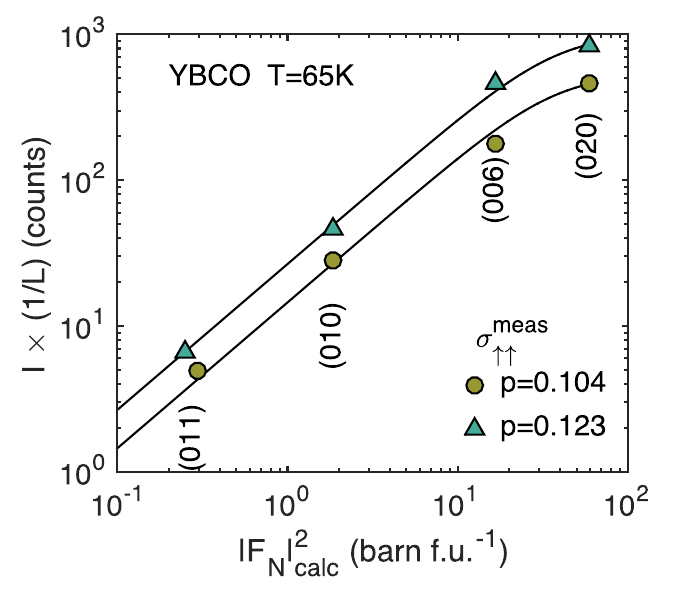}
\caption{A comparison of the measured integrated intensity of $\omega$ (A3) scans of the nuclear Bragg peaks (for $p=0.108,0.123$ samples) times the Lorentz correction factor with the calculated structure factors ($|F_N|^2_{\textrm{calc}}$). The Lorentz factor has been computed using the a full model\cite{Popovici1975_Popo,ResCal2009} for the resolution function of a three-axis spectrometer. The solid lines are fits to a extinction model\cite{Zachariasen1967_Zach,Schultz1986_Schu} which is linear in the limit $|F_N|^2_{\textrm{calc}} \rightarrow 0$.}
\label{I_vs_F_N_log}
\end{figure}

In Fig.~\ref{I_vs_F_N_log} we compare our the measured Bragg intensities, corrected by a full resolution model for the spectrometer, with the calculated values.  The structure factors measured in our experiment are consistent with those predicted by the published structure of YBa$_2$Cu$_3$O$_{6+x}$ and used in our analysis. 
\begin{table}[htb]
\begin{ruledtabular}
\begin{tabular}{l|lllll}
(hkl) &  $I \times (1/L) $(meas) & $|F_N|^2$(calc) & $|F_N|^2$(fit) \\
      &   (arb. units) & (barn f.u.$^{-1}$) & (barn f.u.$^{-1}$) \\ \hline
& \multicolumn{3}{c}{$p=0.104$} \\ 
(011)  &  4.85 $\pm$ 0.18 & 0.28 & 0.33  \\ 
(010)  &  28.6 $\pm$ 0.6 & 1.85 & 1.96  \\ 
(006)  &  178 $\pm$ 5 & 16.7 & 13.0  \\ 
(020)  &  469 $\pm$ 9 & 59 & 66  \\
& \multicolumn{3}{c}{$p=0.123$} \\ 
(011)  &  6.6 $\pm$ 0.2 & 0.20 & 0.24  \\ 
(010)  &  46.8 $\pm$ 0.7 & 1.85 & 1.72  \\ 
(006)  &  468 $\pm$ 9 & 16.7 & 19.5  \\ 
(020)  &  839 $\pm$ 12 & 59 & 56  \\ 
\end{tabular}
\end{ruledtabular}
\caption{\label{Tbl:F_N} Measured integrated intensity of the nuclear Bragg peaks corrected for instrumental resolution compared with the calculated structure factors ($|F_N|^2$). Data from Fig.~\ref{I_vs_F_N_log}.}
\end{table}

\subsection{Alignment Scan}
\begin{figure}[tb]
\includegraphics[width=0.8\linewidth]{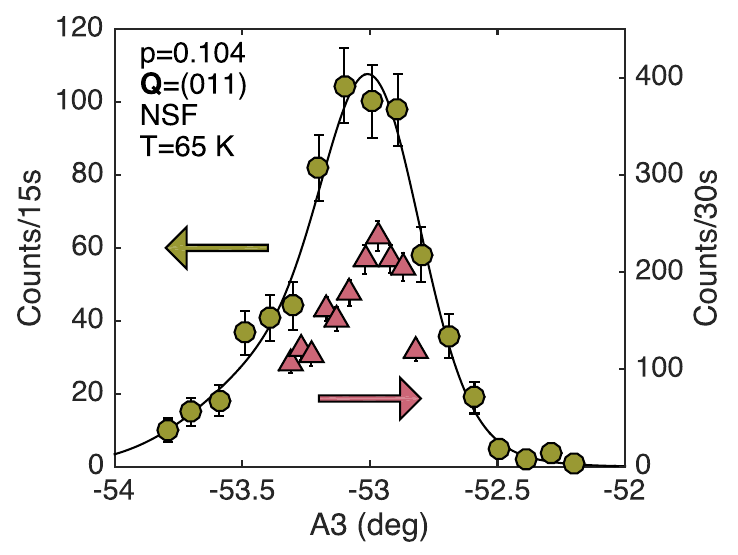}
\caption{Typical alignment scans used to locate the peak in the NSF intensity at the (011) Bragg peak at each temperature.  A broad scan of 0.1$^{\circ}$ steps (circles, left scale) is used to locate the peak, followed by a second scan with 0.05$^{\circ}$ steps (triangles, right scale) over the peak region.}
\label{Bragg011}
\end{figure}

In Sec.~III of the supplementary material of the comment\cite{Bourges2018_BoSM}, our data collection method is questioned.  In Fig.~\ref{Bragg011} we show typical A3 alignment scans performed at each temperature which are used to locate the peak in the NSF intensity as described in Sec.~IIIC of Croft \textit{et al.} \cite{Croft2017_CBKL}.   

%
